\documentclass[prl,aps,showpacs,twocolumn,superscriptaddress,nofootinbib]{revtex4-1}
\usepackage{mathrsfs}
\usepackage{amssymb}
\usepackage{amsmath}
\usepackage{graphicx}

\usepackage[caption = false]{subfig}
\usepackage[normalem]{ulem}
\usepackage{xcolor}
\definecolor{lcolor}{rgb}{0.5,0,0}
\definecolor{citcolor}{rgb}{0,0.3,0.0}

\usepackage[breaklinks,colorlinks,urlcolor=blue,citecolor=citcolor,linkcolor=lcolor]{hyperref}
\usepackage{mciteplus}
\usepackage[utf8]{inputenc}

\newcommand{\pt}{{\mathbf{p}_T}}
\newcommand{\ptt}{|\mathbf{p}_T|}
\newcommand{\abspt}{{p_T}}
\newcommand{\xt}{\mathbf{x}_T}
\newcommand{\absxt}{{x_T}}
\newcommand{\st}{\mathbf{s}}

\newcommand{\n}{\nonumber \\}

\newcommand{\td}{\text{d}}

\newcommand{\dd}{\mathrm{d}}

\usepackage{slashed}
\usepackage{tikz}

\makeindex

\begin{document}

\title{Importance of initial and final state effects for azimuthal correlations in p+Pb collisions}

\author{Moritz Greif}
\email{greif@th.physik.uni-frankfurt.de}
\affiliation{Institut f\"ur Theoretische Physik, Johann Wolfgang Goethe-Universit\"at,
Max-von-Laue-Str.\ 1, D-60438 Frankfurt am Main, Germany}

\author{Carsten Greiner}
\affiliation{Institut f\"ur Theoretische Physik, Johann Wolfgang Goethe-Universit\"at,
Max-von-Laue-Str.\ 1, D-60438 Frankfurt am Main, Germany}

\author{Bj\"orn Schenke}
\affiliation{Physics Department, Brookhaven National Laboratory, Upton, NY 11973, USA}

\author{S\"oren Schlichting}
\affiliation{Department of Physics, University of Washington, Seattle, WA 98195-1560, USA}

\author{Zhe Xu}
\affiliation{Department of Physics, Tsinghua University and Collaborative Innovation Center of Quantum Matter, Beijing 100084, China}

\date{\today}

%%%%%%%%%%%%%%%%%%%%%%%%%%%%%%%%%%%%%%%%%%%%%%%%%%%%%%%%%%%%%%
%%%%%%%%%%%%%%%%%%%%%%%%%%%%%%%%%%%%%%%%%%%%%%%%%%%%%%%%%%%%%%
% A B S T R A C T
%%%%%%%%%%%%%%%%%%%%%%%%%%%%%%%%%%%%%%%%%%%%%%%%%%%%%%%%%%%%%%
%%%%%%%%%%%%%%%%%%%%%%%%%%%%%%%%%%%%%%%%%%%%%%%%%%%%%%%%%%%%%%

\begin{abstract}
We investigate the relative importance of initial and final state effects on azimuthal correlations of gluons in low and high multiplicity p+Pb collisions. To achieve this, we couple Yang-Mills dynamics of pre-equilibrium gluon fields (IP-GLASMA) to a perturbative QCD based parton cascade for the final state evolution (BAMPS) on an event-by-event basis. We find that signatures of both the initial state correlations and final state interactions are seen in azimuthal correlation observables, such as $v_2\left\lbrace2PC\right\rbrace(p_T)$, their strength depending on the event multiplicity and transverse momentum.
Initial state correlations dominate $v_2\left\lbrace2PC\right\rbrace(p_T)$ in low multiplicity events for transverse momenta $p_T>2~{\rm GeV}$. While final state interactions are dominant in high multiplicity events, initial state correlations affect  $v_2\left\lbrace2PC\right\rbrace(p_T)$ for $p_T>2~{\rm GeV}$ as well as the pT integrated $v_2\left\lbrace2PC\right\rbrace$.
	 
\end{abstract}

\maketitle

%%%%%%%%%%%%%%%%%%%%%%%%%%%%%%%%%%%%%%%%%%%%%%%%%%%%%%%%%%%%%%
%%%%%%%%%%%%%%%%%%%%%%%%%%%%%%%%%%%%%%%%%%%%%%%%%%%%%%%%%%%%%%
% I N T R O D U C T I O N
%%%%%%%%%%%%%%%%%%%%%%%%%%%%%%%%%%%%%%%%%%%%%%%%%%%%%%%%%%%%%%
%%%%%%%%%%%%%%%%%%%%%%%%%%%%%%%%%%%%%%%%%%%%%%%%%%%%%%%%%%%%%%

\paragraph*{Introduction.}
\label{sec:Intro}
The measured azimuthal momentum anisotropies of produced particles in heavy ion collisions are well described in the framework of event-by-event hydrodynamics. 
In this picture a fluctuating initial geometry, dominated by fluctuating nucleon positions in the incoming nuclei, is converted into anisotropic momentum space distributions by the pressure driven final state evolution. Hydrodynamic simulations agree well with a wide range of experimental observables from the Relativistic Heavy Ion Collider (RHIC) at Brookhaven National Laboratory and the Large Hadron Collider (LHC) at CERN \cite{Heinz:2013th,Gale:2013da,deSouza:2015ena,Song:2017wtw}.

Measurements in smaller collision systems such as p+p and p+A \cite{Dusling:2015gta}, in particular those of anisotropies in multi-particle correlation functions, have shown very similar features as those in heavy ion collisions. While calculations within the hydrodynamic framework have been quite
successful in describing observables in these small collision systems,
alternative explanations relying entirely on intrinsic momentum
correlations of the produced particles can also reproduce many features
of the experimental data. This includes two and more particle azimuthal correlations and their $p_T$ dependence \cite{Dusling:2015gta,Schlichting:2016sqo,Dusling:2017aot} and mass splitting of identified particle $v_n$ \cite{Schenke:2016lrs}.
Apart from the existence of alternative explanations, the applicability of hydrodynamics becomes increasingly doubtful as the system size decreases and gradients increase. Some recent studies argue that hydrodynamics should be applicable in systems of sizes down to $\sim 0.15\,{\rm fm}$ \cite{Romatschke:2016hle}, but off-equilibrium corrections to particle distribution functions for momenta $p_T\gtrsim 0.5\,{\rm GeV}$ can be significant \cite{Mantysaari:2017cni}, which limits at least the quantitative reliability of the framework.

So far all calculations of multi-particle correlations in small collision systems have studied either only intrinsic momentum correlations or purely final state driven effects. 
In this letter we  present the first study where both effects are combined into a single framework to assess their relative importance. 

We compute initial state gluon Wigner-distributions from the Impact Parameter dependent Glasma model (IP-Glasma) \cite{Schenke:2012wb,Schenke:2012hg} and via sampling of individual gluons feed them into the partonic transport simulation 'Boltzmann approach to multiparton scatterings' (BAMPS) \cite{Xu:2004mz}. The initial gluon distributions \cite{Krasnitz:1999wc,Krasnitz:2000gz} from the IP-Glasma model are anisotropic in momentum space \cite{Lappi:2009xa,Schenke:2015aqa,Schenke:2016lrs}, thus contain the intrinsic momentum space correlations of the color glass condensate (CGC)  picture \cite{Gelis:2010nm,Gelis:2016upa}. Final state interactions mediated by perturbative quantum chromo dynamic (pQCD) cross sections are then simulated microscopically in BAMPS. 

We analyze the time evolution of the momentum space anisotropy of the partonic plasma by simulating events in two different multiplicity classes to understand how final state interactions modify initial state momentum correlations and whether signals of the latter can survive to affect final observables. 

\paragraph*{Initial state \& Phase-space distribution.}
\label{sec:InitialState}
Based on the IP-Glasma model, including event-by-event fluctuations of the proton's geometrical structure\,\cite{Mantysaari:2016ykx}, we calculate the solution to the classical Yang-Mills equations of motion up to $\tau_0=0.2\,{\rm fm}/c$ following the standard procedures described in \cite{Schenke:2012wb,Schenke:2012hg}. Event-by-event we extract the Wigner distribution $\frac{\dd N_{g}}{\dd y \dd^2\xt \dd^2\pt}$ in hyperbolic phase-space coordinates $x^\mu=(\tau\cosh \eta_s,\xt,\tau\sinh \eta_s),\;$ $p^\mu=(\ptt\cosh y,\pt,\ptt\sinh y),\;$ by evaluating equal time correlation functions in Coulomb gauge and projecting them onto the transversely polarized mode functions $\xi^{(\lambda)}_{\pt}(\tau)$ of the free theory \cite{Berges:2013fga}, according to 
\begin{align}
\label{eq:CYMDistribution}
\frac{\dd N_{g}}{\dd y \dd^2\xt \dd^2\pt}&=\frac{1}{(2\pi)^2} \sum_{\lambda=1,2} \sum_{a=1}^{N_c^2-1}  \tau^2~g^{\mu\mu'} g^{\nu\nu'} \nonumber \\
\times\int \dd^2\st~ &\Big(\xi^{(\lambda)*}_{\pt,\mu}(\tau) i\overleftrightarrow{\partial_{\tau}}   A_{\mu'}^{a}(\xt+\st/2) \Big)  \nonumber \\
&\Big(A_{\nu'}^{a}(\xt-\st/2)  i\overleftrightarrow{\partial_{\tau}} \xi^{(\lambda)}_{\pt,\nu}(\tau) \Big)    e^{-i\pt\cdot \st}\;.
\end{align}
Even though the position and momentum dependent Wigner distribution includes all relevant information about the initial state coordinate space eccentricity as well as the initial state momentum space anisotropies, it suffers from the deficiency that it is not necessarily positive semi-definite. To warrant a probabilistic interpretation of a quasi-particle distribution entering the subsequent Boltzmann transport simulation, it is necessary to perform a smearing of the Wigner distribution over phase space volumes $\sigma_{x} \sigma_{p} \geq \hbar/2$. Accounting for the boost-invariant nature of the classical Yang-Mills fields the single particle distribution function $f^{g}_{0}$, which will enter the subsequent parton cascade, is obtained by performing the Gaussian smearing 
\begin{eqnarray}
\label{eq:InitialPhaseSpaceDensity}
&&f^{g}_{0}(\xt,\eta_s,\mathbf{p}_\perp,y)=\frac{(2\pi)^3}{2(N_c^2-1)}  \frac{\delta(y-\eta_s)}{\ptt\tau}  \\
&&\qquad \times \int \frac{\dd^2\xt' \dd^2\pt'}{(2\pi)^2} e^{-\frac{(\xt-\xt')^2}{2\sigma_x^2}} e^{-\frac{(\pt-\pt')^2}{2\sigma_p^2}} \frac{\td N_{g}}{\td y \td^2\xt'\td^2\pt'}\;, \nonumber
\end{eqnarray}
with $\sigma_x=0.197~\rm{fm}$ and $\sigma_p=1~\rm{GeV}$ chosen to achieve a reasonable compromise between spatial and momentum resolution.\\

\paragraph*{Final state interactions.}
\label{sec:BAMPS}
Even though the classical Yang-Mills evolution includes re-scattering effects at early times, the semi-classical description of the dynamics becomes inapplicable after a relatively short time when quantum effects become important and the subsequent dynamics is more appropriately described in terms of weakly interacting quasi-particles \cite{Baier:2000sb,Berges:2013fga,Kurkela:2015qoa}. We simulate the dynamics within $0.2~\rm{fm/c} < \tau < 2.0~\rm{fm/c}$, with a 3+1-dimensional Boltzmann approach to multi-parton scatterings (BAMPS), which, starting from the initial phase-space density of gluons in Eq.~(\ref{eq:InitialPhaseSpaceDensity}), solves the relativistic Boltzmann equation
\begin{equation}
p^{\mu }\frac{\partial }{\partial x^{\mu }}f^i(x,p)=\sum\limits_{j=g,q,\overline{q}} C_{ij}(x,p),
\label{eq:BE}
\end{equation}%
for the phase-space distribution function $f^i(x,p)$ of massless on-shell quarks, anti-quarks and gluons by Monte-Carlo techniques \cite{Xu:2004mz,Xu:2007aa,Xu:2014ega}\footnote{Even though the IP-Glasma initial state only contains gluons, quarks and anti-quarks are produced during the kinetic evolution of the fireball.}. The collision integrals $C_{ij}$ include $2\leftrightarrow2$ and $2\leftrightarrow3$ interactions, based on perturbative QCD matrix elements (using a fixed strong coupling constant $\alpha_s=0.3$) where internal propagators are regulated by a dynamically computed screening mass $m_D^2\sim \alpha_s\int d^3p f^i(x,p)/p$ (see, e.g., Refs.~\cite{Fochler:2013epa,Uphoff:2014cba}). Inelastic $2\leftrightarrow3$ interactions are simulated based on the improved Gunion-Bertsch matrix elements \cite{Fochler:2013epa}, and the Landau-Pomeranchuk-Migdal (LPM) effect is treated effectively, based on a dynamically determined mean free path \cite{Uphoff:2014cba}.

Since in practice the Monte-Carlo implementation is based on individual particles, propagating along straight lines between scattering events, one needs to supply a list of particle positions $x^{\mu}_{\rm Init}$ and momenta $p^{\mu}_{\rm Init}$ as initial condition for BAMPS. For every event we sample a collection of individual gluons from the momentum distribution $f^{g}_{0}(\xt,\eta_s,\pt,y)$ of the IP-Glasma model, such that the overall number of gluons is given by the integral of the distribution. Since according to Eq.~(\ref{eq:InitialPhaseSpaceDensity}) the initial momentum rapidity $y$ is equal to the coordinate space rapidity $\eta_s$, which we sample uniformly between $-2<\eta_s<2$ from the boost invariant distribution, the initial position and momentum vectors of each particle are given by $x^{\mu}_{\rm Init}=\left(\tau_0\cosh(\eta_s),\xt,\tau_0\sinh(\eta_s)\right)$ and $p^{\mu}_{\rm Init}=\left(\ptt\cosh(\eta_s),\pt,\ptt\sinh(\eta_s)\right)$. 

We have checked explicitly, that the energy density $(T^{\tau\tau})$ and flow coefficients $(v_2)$ extracted from the sampled particle ensemble agree well with the corresponding quantities extracted directly from the IP-Glasma distribution. Even though the IP-Glasma initial condition is boost invariant, the BAMPS calculation is performed in 3+1 dimensional Minkowski space. We will therefore extract all observables at $|y|<0.5$ for different lab times $t$, where $y=\log[(E+p_z)/(E-p_z)]/2$, noting that at midrapidity $|y|\approx |\eta_s| \approx 0$ such that the lab time  $t\approx \tau$.\\

\paragraph*{Evolution of azimuthal anisotropies.}
We investigate the evolution of the azimuthal momentum space anisotropy characterized by the Fourier harmonics $v_{n}\{2PC\}$ of the two-particle correlation function. We follow the experimental analysis\,\cite{Chatrchyan:2013nka} in decomposing the (normalized) two-particle correlation function 
%$\frac{2\pi}{N_\text{trig}N_\text{assoc}}\frac{\td N^\text{pair}}{\td\Delta\varphi}(p_T,p_T^{\text{ref}})$, 
for $N_\text{trig}$ trigger particles in a momentum range given by $p_T^{\rm ref}$ and $N_\text{assoc}$ particles in a momentum bin around $p_T$, in Fourier harmonics w.r.t. the relative azimuthal angle $\Delta\varphi_\abspt$:
\begin{align}
\frac{2\pi}{N_\text{trig}N_\text{assoc}}\frac{\td N^\text{pair}}{\td\Delta\varphi_\abspt}&(p_T,p_T^{\text{ref}})=\n
1+\sum\limits_n &2V_{n\Delta}(p_T,p_T^{\text{ref}})\cos(n\Delta \varphi_\abspt).
\end{align}
The two particle $v_2\{2PC\}$ is obtained as \cite{Chatrchyan:2013nka}
\begin{align}
v_n\{2PC\}(p_T)=\frac{V_{n\Delta}(p_T,p_T^{\text{ref}})}{\sqrt{V_{n\Delta}(p_T^{\text{ref}},p_T^{\text{ref}})}},
\end{align} 
with the reference momentum range chosen as $0~\mathrm{GeV}<p_T^{\text{ref}} < 8~\mathrm{GeV}$ by default\footnote{Because we are studying the momentum anisotropy of gluons, we choose the reference momentum to extend to larger values than the range used in the experimental analysis.}. Since in our model the double-inclusive spectrum in each event is given by the product of single inclusive spectra, we follow \cite{Lappi:2015vha,Schenke:2015aqa} and directly compute 
\begin{eqnarray}\label{eq:bn}
V_{n\Delta}(p_T,p_T^{\text{ref}})=\left\langle  \text{Re} \frac{b_{n}(p_T) b_{n}^{*}(p_T^{\text{ref}})}{b_{0}(p_T) b_{0}^{*}(p_T^{\text{ref}})} \right\rangle_{\rm events}
\end{eqnarray}
where in each event $b_{n}(p_T)=\int \frac{\dd\phi_{\abspt}}{2\pi} \frac{\dd N_g}{\dd^2\pt} e^{in\phi_{\abspt}} $ is the azimuthal Fourier coefficient of the single-inclusive spectrum. Since our model does not include correlations from back-to-back di-jet pairs, we also note that  -- contrary to the experimental analysis -- no additional subtractions are required to eliminate such correlations.\\ 

\begin{figure}
\centering
\includegraphics[width=\columnwidth]{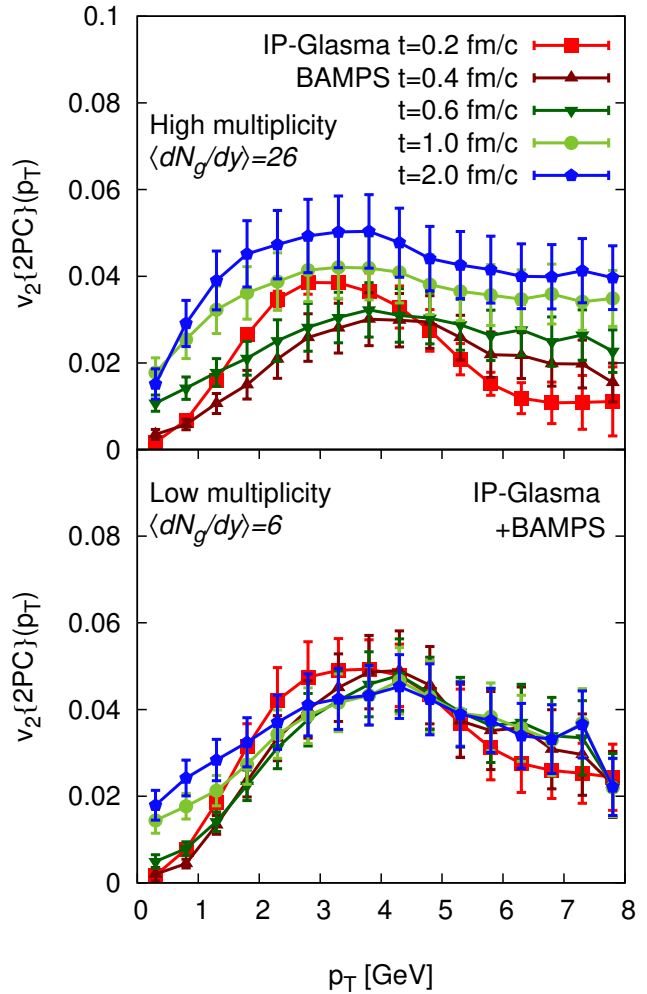}
\caption{Gluon $v_2\{2PC\}(\abspt)$ at mid-rapidity ($|y|<0.5$) for different times in high multiplicity ($\langle dN_{g}/dy\rangle=26$, upper panel) and low multiplicity ($\langle dN_{g}/dy\rangle=6$, lower panel) p+Pb collisions.}
\label{fig:v2_2PC}
\end{figure}

\paragraph{Evolution of azimuthal anisotropy.} Including both initial state effects and final state evolution, we analyze the time evolution of the momentum space anisotropy $v_2\{2PC\}(p_T)$ for $\sqrt{s_{\rm pA}}=5.02\,{\rm TeV}$ p+Pb collisions in Fig.\,\ref{fig:v2_2PC}. We show $v_2\{2PC\}(p_T)$ at different times, $t=0.2\, \text{(initial)},0.4,0.6,1,2\,\mathrm{fm}/c$ for low multiplicity ($0.5 < \left( \dd N_g/\dd y\right)/ \langle \dd N_g/\dd y \rangle  < 1$) and high multiplicity ($\left( \dd N_g/\dd y\right)/ \langle \dd N_g/\dd y \rangle  > 2.5$) events.

While in both cases momentum correlations lead to a sizeable initial state $v_{2}$ \cite{Schenke:2015aqa}, the subsequent dynamics is quite different: In high multiplicity events, we observe a pronounced effect of the final state interactions such that the high initial anisotropy at intermediate momenta $(\abspt\sim2-5~\mathrm{GeV})$ is significantly reduced within the first $0.2~\mathrm{fm/c}$ evolution in the parton cascade, while at the same time the correlation strength at higher and lower momenta begins to increase. Subsequently, the azimuthal anisotropy increases for all $\abspt$ up to maximally $5\%$. As a result, the pronounced peak at around $\abspt\sim 3~\mathrm{GeV}$, present after the IP-Glasma stage, is washed out by the final state interactions. In contrast, for low multiplicity events modifications due to final state effects appear to be less significant, as the final curve $v_{2}(p_T)$ closely resembles that of the IP-Glasma initial state. Only at low transverse momenta, $\abspt\lesssim 2~\mathrm{GeV}$ the azimuthal anisotropy is increased to $2-3~\%$. 

While our results confirm the basic expectation that final state effects gain significance as the density of the medium increases in high-multiplicity events~\cite{Schlichting:2016xmj,Schlichting:2016sqo}, the way this is realized dynamically is in fact very interesting. We find that the average number of interactions in low-multiplicity events ($N_{\rm{scat}}=4.5 \pm 1.1$) is indeed almost the same as in high-multiplicity events ($N_{\rm{scat}}=5.6 \pm 1.1$). Because of the nature of the QCD cross-sections, most interactions however correspond to small momentum transfers ${\sim}m_D$ which itself depends on the density of the medium~\cite{Arnold:2002zm}, such that the average momentum transfer is larger in high-multiplicity events. Hence, the average number of \emph{large angle scatterings}, estimated according to $N_{\rm{scat}}^{\rm{large~angle}}= \frac{1}{N_{\rm{particles}}} \sum_{\rm{coll}} \frac{3}{2} \sin^2 \theta^{\rm{coll}}_{\rm{c.o.m.}}$ where $\theta^{\rm{coll}}_{\rm{c.o.m.}}$ is the scattering angle in the c.o.m. frame of the partonic interaction\footnote{Note that the pre-factor $3/2$ is chosen such that for constant isotropic cross sections $N_{\rm{scat}}^{\rm{large~angle}}= N_{\rm{scat}}$.},  is in fact significantly larger in high-multiplicity events $(N_{\rm{scat}}^{\rm{large~angle}}=1\pm 0.18)$ as compared to low-multiplicity events $(N_{\rm{scat}}^{\rm{large~angle}}=0.53 \pm 0.14)$.\\

\begin{figure}
 	\centering
 	\includegraphics[width=\columnwidth]{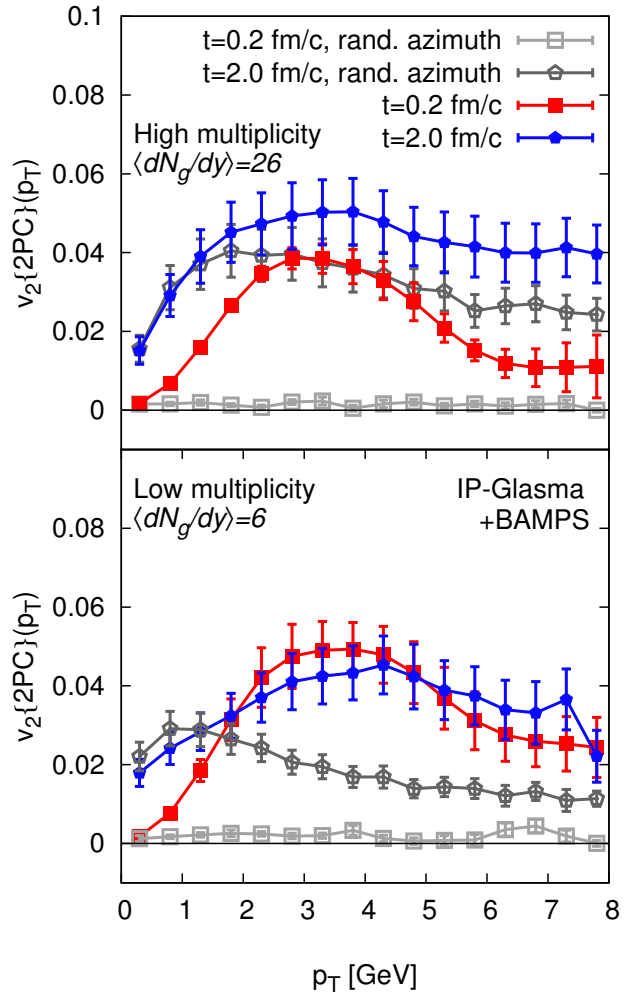}
 	\caption{Comparison of initial and final two-particle $v_2(p_T)$ for high (upper panel) and low (lower panel) multiplicity $\sqrt{s_{pA}}=5.02~\mathrm{TeV}$ p+Pb events. Events including initial state momentum correlations (filled symbols) are compared to the same events where the initial momenta were randomized in azimuth (rand. azimuth, open symbols).}
 	\label{fig:RandAzimuthLowHigh}
 \end{figure}

\paragraph*{Initial state vs. final state effects.}
In order to further disentangle the effects of initial state momentum correlations and final state response to geometry, we performed an additional set of simulations (henceforth labeled rand. azimuth) where the azimuthal angle of the transverse momentum $\pt$ of each gluon is randomized ($0<\varphi_\abspt<2\pi$) before the evolution in the parton cascade. Our results are compactly summarized in Fig.~\ref{fig:RandAzimuthLowHigh}, where we compare the azimuthal anisotropy $v_2\{2PC\}(p_T)$ in the different scenarios. By construction no initial state momentum correlations are present in the rand. azimuth case -- shown as open gray symbols -- and the initial state $v_2$ vanishes identically at $t=0.2~\mathrm{fm/c}$. However, over the course of the kinetic evolution a $v_{2}(p_T)$ of $\sim4\%$ at $\abspt\sim2~\mathrm{GeV}$ in high multiplicity events and $\lesssim 3\%$ at $\abspt~\sim 1~\mathrm{GeV}$ in low multiplicity events is built up by $t=2.0~\mathrm{fm/c}$. Nevertheless, for momenta above $\abspt\sim 2.0~\mathrm{GeV}$ (low multiplicity) and $\abspt\sim 4.0~\mathrm{GeV}$ (high multiplicity), the purely final state $v_2$ in the rand. azimuth scenario remains significantly below the initial state + final state $v_{2}$ of the full calculation, indicating the importance of initial state momentum correlations. 

\begin{figure}
\centering
\includegraphics[width=\columnwidth]{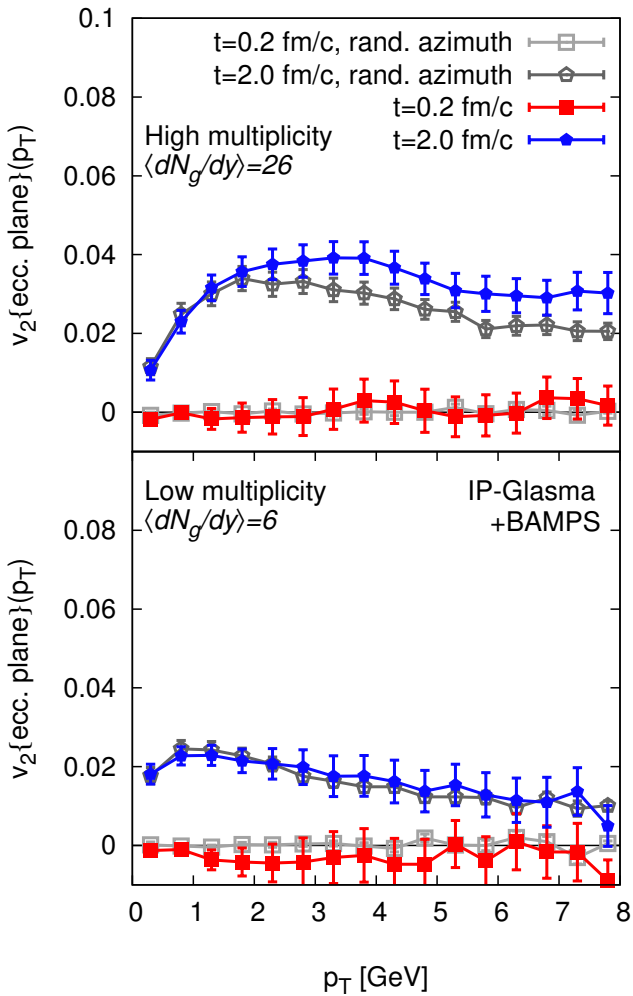}
\caption{Comparison of initial and final $v_2(p_T)$ with respect to the eccentricity plane for high (upper panel) and low (lower panel) multiplicity $\sqrt{s_{pA}}=5.02~\mathrm{TeV}$ p+Pb events. Events including initial state momentum correlations (filled symbols) are compared to the same events where the initial momenta were randomized in azimuth (rand. azimuth, open symbols).}
\label{fig:RandAzimuthLowHigh_EccentrictiyPlane}
\end{figure}

Despite the fact that initial state correlations have a significant impact on $v_2\{2PC\}$, we find that the additional $v_2\{2PC\}$ built up in the parton cascade can be attributed to the response to the initial geometry. In order to demonstrate this feature more clearly, we have also computed the azimuthal anisotropy $v_2\{\text{ecc.~plane}\}$ w.r.t to the coordinate eccentricity plane -- obtained by replacing the reference momentum vector $b_{n}(p_T^{\text{ref}})$ in Eq.~(\ref{eq:bn}) with the coordinate eccentricity vector $e_{n}=\int d^2\xt~T^{\tau\tau}(\xt)~|\xt|^{n}~e^{in \phi_{\absxt}}$, where $\phi_{\absxt}$ is the azimuthal angle in space. Our results in Fig.~\ref{fig:RandAzimuthLowHigh_EccentrictiyPlane} show that the initial anisotropy with respect to the geometric eccentricity plane vanishes, as the initial momentum space anisotropy is uncorrelated with the event geometry \cite{Schenke:2015aqa}.

In contrast, during the kinetic evolution a clear correlation with the initial state geometry is built up. The magnitude of this final state generated $v_2\{\text{ecc. plane}\}$ depends only weakly on the presence or absence of initial state momentum correlations. While the comparison of the results for $v_2\{\text{ecc. plane}\}$ (Fig.\,\ref{fig:RandAzimuthLowHigh_EccentrictiyPlane}) with $v_{2}\{2PC\}$ (Fig.\,\ref{fig:RandAzimuthLowHigh}) indicates that in the rand. azimuth case, the observed  $v_{2}\{2PC\}$ can almost entirely be attributed to a geometric response, this is clearly not the case for the more realistic scenario including initial state correlations.

\begin{figure}
	\centering
	\includegraphics[width=\columnwidth]{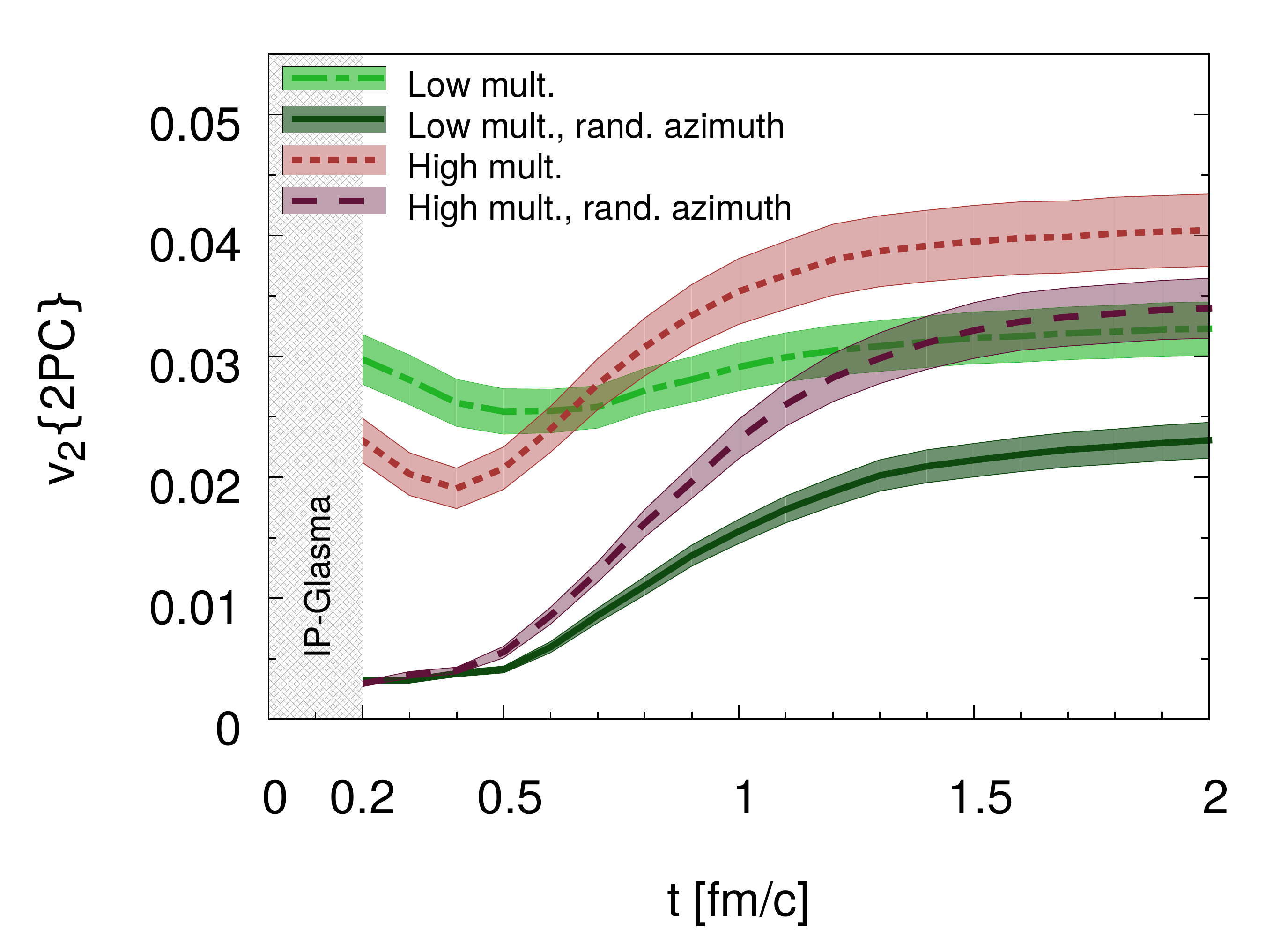}
	\caption{Evolution of the $\abspt$ integrated azimuthal anisotropy $v_{2}\{2PC\}$ for high and low multiplicity p+Pb events.}
	\label{fig:integratedV2}
\end{figure}

Even though the effects of initial state momentum correlations are more apparent in low-multiplicity events, quantitative differences remain also in high-multiplicity events, as can also be observed from Fig.~\ref{fig:integratedV2}, where we study the time-evolution of the $\abspt$ integrated $v_2\{2PC\}$. While in the rand. azimuth case, the $v_2\{2PC\}$ is built up slowly as a function of time in response to the initial state geometry, a qualitatively different behavior emerges in the more realistic case including initial state correlations. In this case, large angle scatterings at early times begin to destroy initial state momentum correlations leading to an initial decrease of $v_2\{2PC\}$ as a function of time. This happens because the directions of the initial state anisotropy and the eccentricity responsible for generating the final state $v_2$ are generally uncorrelated. Subsequently, between $t\sim 0.5 - 1~\mathrm{fm/c}$ the response to the initial state geometry sets in, leading again to an increase of $v_2\{2PC\}$. Overall, we find that the relative effect of initial state correlations on the final $v_2\{2PC\}$ is on the order of $25-50\%$, being larger for low multiplicity events.

\paragraph*{Conclusions.}
\label{sec:Conclusions}
The observation of long range rapidity correlations
with characteristic structures in azimuthal angle in small systems has challenged
our understanding of the space-time evolution of high-energy nuclear collisions.
Despite the fact that several phenomenological works have attempted to explain various aspects of the experimental data, it remained unclear to what extent observed correlations should be attributed to initial state or final state effects. Based on a weak-coupling picture of the space-time dynamics, we developed a new framework including both initial state momentum correlations and final state interactions. By matching classical Yang-Mills dynamics (IP-GLASMA) to an effective kinetic description (BAMPS) on an event-by-event basis, we showed that the relative importance of initial and final state effects in p+Pb collisions at LHC energies depends on the event multiplicity as well as the transverse momenta under consideration. 
Especially at low multiplicity, the initial state correlations are very important for integrated as well as differential $v_{2}$, and need
to be taken into account in a quantitative theoretical description.

We also note that multi-particle correlations of more than two particles can provide additional insight into the nature of the observed correlations. Since final state induced correlations emerge in response to the global event-geometry, these naturally produce $m$-particle correlations (with $m>2$) of similar strength. Conversely, for initial state correlations the existence of pronounced multi-particle correlations is not a priori obvious. However, it was shown recently in an Abelian model that initial state effects can generate similar 4-, 6-, and 8- particle correlations \cite{Dusling:2017dqg}. Explicit studies of multi-particle correlations beyond $m=2$ within our framework are numerically very intensive and will be left for future work.
Our results indicate that a differential study of azimuthal correlations across a large
range of multiplicities and transverse momenta, can provide
new insights into properties of the initial state and the
early time non-equilibrium dynamics of high-energy collisions. In this context, it would also be interesting to include jet-like correlations at higher momenta, to achieve a fully comprehensive framework of multi-particle correlations.

\paragraph*{Acknowledgements.}
M.G. is grateful to Tsinghua University in Beijing for their hospitality and acknowledges the support from the ``Helmhotz Graduate School for Heavy Ion research''. This work was supported by the Helmholtz International Center for FAIR within the framework of the LOEWE program launched by the State of Hesse. S.S. acknowledges support by the U.S. Department of Energy (DOE) under Grant No. DE-FG02-97ER41014. B.P.S. is supported under DOE Contract No. DE-SC0012704. Z.X. was supported by the National Natural Science Foundation of China under Grants No. 11575092 and No. 11335005, and the Major State Basic Research Development Program in China under Grants No. 2014CB845400 and No. 2015CB856903. Numerical calculations used the resources of the Center for Scientific Computing (CSC) Frankfurt and the National Energy Research Scientific Computing Center, a DOE Office of Science User Facility supported by the Office of Science of the U.S. Department of Energy under Contract No. DE-AC02-05CH11231.

%%%%%%%%%%%%%%%%%%%%%%%%%%%%%%%%%%%%%%%%%%%%%%%%%%%%%%%%%%%%%
%%%%%%%%%%%%%%%%%%%%%%%%%%%%%%%%%%%%%%%%%%%%%%%%%%%%%%%%%%%%%
% B I B L I O G R A P H Y
%%%%%%%%%%%%%%%%%%%%%%%%%%%%%%%%%%%%%%%%%%%%%%%%%%%%%%%%%%%%%
%%%%%%%%%%%%%%%%%%%%%%%%%%%%%%%%%%%%%%%%%%%%%%%%%%%%%%%%%%%%%

\bibliographystyle{apsrev4-1}
\bibliography{library_manuell.bib}

\end{document}